\documentclass[superscriptaddress,aps, prd, reprint, nofootinbib]{revtex4-2}
\pdfoutput=1
\usepackage{amsmath}
\usepackage{amssymb}
\usepackage{amsbsy, amstext, amsthm}
\usepackage{dcolumn}% Align table columns on decimal point
\usepackage{bm}% bold math
\usepackage{blkarray}
\usepackage{slashed}          
\usepackage{mathtools}                   
\usepackage{hyperref,graphicx}           % usepackage without driver option
\usepackage{url}
\usepackage{caption}
\usepackage{subcaption}
\usepackage{multirow}
\usepackage{units}
\usepackage{physics}
\usepackage[usenames,dvipsnames,svgnames,table]{xcolor}
\usepackage{graphicx}
\usepackage{tikz}
\usepackage{siunitx}[=v2]
\usepackage[all]{hypcap}
\usepackage{cleveref}
\usepackage[mathlines]{lineno}
\usepackage{tikz-3dplot}
\usepackage{color}
\usepackage{multirow}
\usepackage{float}

\numberwithin{equation}{section}
\definecolor{darkblue}{rgb}{0,0,0.5}
\definecolor{darkgreen}{rgb}{0,0.5,0}

\newcommand{\cmt}[1]{}

\newcommand{\MPl}{M_\text{Pl}}

\let\deltafunc\delta
\DeclareDocumentCommand\delta{}{\trigbraces{\deltafunc}}

\graphicspath{{./Figure/}}

\begin{document}
\title{Opening up a Window on the Postinflationary QCD Axion}
\date{June 12, 2023}

\author{Yunjia Bao}
\email{yunjia.bao@uchicago.edu}
\affiliation{Department of Physics, University of Chicago, Chicago, IL, 60637, USA}

\author{JiJi Fan}
\email{jiji\_fan@brown.edu}
\affiliation{Department of Physics \& Brown Theoretical Physics Center, Brown University, Providence, RI, 02912, USA}

\author{Lingfeng Li}
\email{lingfeng\_li@brown.edu}
\affiliation{Department of Physics, Brown University, Providence, RI, 02912, USA}

\begin{abstract}
The QCD axion cosmology depends crucially on whether the QCD axion is present during inflation or not. We point out that contrary to the standard criterion, the Peccei-Quinn (PQ) symmetry could remain unbroken during inflation, even when the axion decay constant, $f_a$, is (much) above the inflationary Hubble scale, $H_I$. This is achieved through the heavy-lifting of the PQ scalar field due to its leading non-renormalizable interaction with the inflaton, encoded in a high-dimensional operator which respects the approximate shift symmetry of the inflaton. The mechanism opens up a new window for the post-inflationary QCD axion and significantly enlarges the parameter space, in which the QCD axion dark matter with $f_a > H_I$ could be compatible with high-scale inflation and free from constraints on axion isocurvature perturbations. There also exist non-derivative couplings, which still keep the inflaton shift symmetry breaking under control, to achieve the heavy-lifting of the PQ field during inflation. 
Additionally, by introducing an early matter domination era, more parameter space of high $f_a$ could yield the observed DM abundance. \end{abstract}

\maketitle

%%%%%%%%%%%%%%%%%%%%%%%%%%%%%%%%%%%%%%%%%%%%%%%
\section{Introduction}
%%%%%%%%%%%%%%%%%%%%%%%%%%%%%%%%%%%%%%%%%%%%%%%
The QCD axion is of great interest in solving the strong CP problem of the standard model~\cite{Peccei:1977hh, Peccei:1977ur,Weinberg:1977ma,Wilczek:1977pj}, and serving as a classical cold dark matter (DM) candidate~\cite{Preskill:1982cy,Dine:1982ah,Abbott:1982af}. It is a naturally light pseudo-Nambu-Goldstone boson from spontaneous breaking of a global symmetry---the $U(1)$ Peccei-Quinn (PQ) symmetry---at a high energy scale $f_a$, which is often referred to as the axion decay constant. It is one of the best motivated feebly-coupled particles beyond the standard model and has attracted continuously growing theoretical and experimental studies even four decades after the original proposal (e.g., discussions of future plans and open questions in some recent Snowmass studies~\cite{Adams:2022pbo, Agrawal:2022yvu, Asadi:2022njl}). 

The QCD axion has a close intriguing interplay with cosmic inflation, a leading paradigm to understand the origin of our universe~\cite{Guth:1980zm, Linde:1981mu,Albrecht:1982mp}. Depending on when the PQ symmetry is broken, the QCD axion cosmologies fall into two classes: the inflationary axion with the PQ symmetry breaking during inflation and the post-inflationary axion (PIA) with the PQ breaking happening afterward. To determine which class the QCD axion is in, we need to compare $f_a$ with the temperature of the universe during inflation, which is given by the Gibbons-Hawking temperature~\cite{Gibbons:1977mu}, $T_{\rm GH} = H_I/(2\pi)$ with $H_I$ the inflationary Hubble scale. More details could be found in recent reviews~\cite{Sikivie:2006ni,Marsh:2015xka, DiLuzio:2020wdo}. 

When $f_a > H_I/(2\pi)$, the QCD axion is in the first category and is present during inflation.\footnote{The assumption here is that PQ symmetry is not restored during the (p)reheating epoch after inflation. This could happen if the maximum thermalization temperature achieved before reheating starts is very high and above $f_a$~\cite{Hertzberg:2008wr}. It is model dependent: e.g., it is not realized in the simple perturbative reheating scenario with inflaton decaying through non-renormalizable operators suppressed by the Planck scale, after taking into account of the maximum thermalization temperature. Thermal restoration could be more likely to happen in more complicated PQ models with light PQ fields~\cite{Kawasaki:2013ae}. Our key new point will be independent of the (p)reheating model, and could be viewed as an alternative way to restore the PQ symmetry.} In this case, the exponential expansion stretches each patch of the initial misalignment angle $\theta_{i}$, and our current Hubble patch owns a single uniform central value of $\theta_i$ at the end of inflation. The massless axion during inflation undergoes random quantum fluctuations set by $T_{\rm GH}$, resulting in an axion isocurvature perturbation $\langle \delta \theta_{i}^2 \rangle=H_I^2/(2 \pi f_a)^2$~\cite{Steinhardt:1983ia, Seckel:1985tj, Lyth:1989pb, Turner:1990uz, Linde:1991km}. This leads to the serious axion isocurvature problem: inflationary QCD axion DM could be incompatible with high-scale inflation (e.g., $f_a \sim 10^{11}$ GeV requires $H_I \lesssim 10^7$ GeV), due to the observation bound on the isocurvature perturbations~\cite{Planck:2018jri}.\footnote{Solutions to the axion isocurvature problem have been proposed. For example, the isocurvature perturbation is suppressed if during inflation, the PQ breaking scale is temporarily enhanced~\cite{Linde:1991km, Choi:2014uaa, Chun:2014xva, Fairbairn:2014zta, Higaki:2014ooa, Nakayama:2015pba, Harigaya:2015hha, Kearney:2016vqw} or QCD is much stronger~\cite{Jeong:2013xta}.}

When $f_a < H_I/(2\pi)$, the QCD axion is in the post-inflation regime, with a quite different cosmic evolution. Due to the absence of axion during inflation, there is no axion isocurvature problem. Each Hubble patch has a $\theta_{i}$ drawn randomly from a uniform distribution between $-\pi$ and $\pi$, with $\langle \theta_{i}^2 \rangle = \pi^2/3$. In addition to the misalignment component, topological defects such as axion strings and domain walls \footnote{If the axion domain wall number is larger than one, the domain walls could be long-lived and incompatible with the observations~\cite{Sikivie:1982qv}. We will assume the domain wall number to be one and thus the walls decay during the early universe~\cite{Vilenkin:1982ks,Lazarides:1982tw,Chang:1998tb}.} are generated, and their decays contribute to the axion abundance~\cite{Davis:1986xc,Vilenkin:1986ku,Harari:1987ht,Davis:1989nj,Battye:1993jv,Battye:1994au,Yamaguchi:1998gx,Hagmann:2000ja,Gorghetto:2020qws,Buschmann:2021sdq}. The latest simulation indicates that the decays of the topological defects could be the dominant contribution and $f_a \sim 10^{10-11}$ GeV for the QCD axion to be the major component of DM in the minimal scenario~\cite{Gorghetto:2020qws,Buschmann:2021sdq}. For this case, ultra-dense compact minihalos might be formed as proposed in~\cite{Hogan:1988mp,Kolb:1993zz, Kolb:1993hw, Kolb:1994fi}, giving rise to a plethora of interesting phenomenology and detection opportunities~\cite{Zurek:2006sy, Vaquero:2018tib, Tinyakov:2015cgg, Davidson:2016uok, Fairbairn:2017sil, Bai:2017feq, Dror:2019twh, Dai:2019lud, Shen:2022ltx}. 

In this article, we will point out that the inclusion of the leading high-dimensional operator between the inflaton and the PQ scalar field (the phase becomes the QCD axion after symmetry breaking) could change the standard story briefly reviewed above significantly. In particular, it will modify the conventional boundary between inflationary and post-inflationary axion drastically. We require the operator to respect the (approximate) shift symmetry of the inflaton.\footnote{Ref.~\cite{Shafi:1984tt,Kofman:1986wm,Kofman:1988xg,Hodges:1989dw,Linde:1991km,Gorghetto:2021fsn} discuss renormalizable interactions between the inflaton and the PQ scalar field, which could lead to large quantum corrections to the inflaton potential and works only in the specific hybrid inflation scenario.} The non-renormalizable operator, though suppressed by a high cutoff scale, could open up a new window for the PIA. More specifically, we will show that even when $f_a \gg H_I/(2\pi)$, the PQ symmetry could remain unbroken during inflation thanks to ``heavy-lifting" of the high-dimensional operator. As a result, part of the region in the $(H_I, f_a)$ plane used to be labeled as inflationary axion and suffer from the axion isocurvature problem could belong to the PIA. The cosmological evolutions and consequences will be modified correspondingly. We will also comment on other types of operators that could achieve the same goal. 

Our mechanism also enlarges the parameter space in which high-scale inflation scenarios and the QCD axion DM could be compatible with each other. QCD axion DM with $f_a > H_I/(2 \pi)$ will not lead to unacceptable isocurvature perturbations since it is not present during inflation in our model. Large $f_a$ may lead to too much QCD axion DM relic abundance through the standard production, which could be diluted subsequently to the observed value by the early matter domination (EMD) mechanism~\cite{Steinhardt:1983ia,Lazarides:1990xp,Kawasaki:1995vt, Banks:1996ea, Giudice:2000ex,Grin:2007yg,Visinelli:2009kt,Bernal:2021yyb,Arias:2021rer,Bernal:2021bbv,Arias:2022qjt}. 

 We present our inflationary heavy-lifting mechanism in Sec.~\ref{sec:heavy-lift} and map out the new window for PIA in Sec.~\ref{sec:newwindow}. We conclude in Sec.~\ref{sec:conc}. Details of the relic abundance computations in the EMD epoch are collected in App.~\ref{app:EMDRelicAbund}.

%%%%%%%%%%%%%%%%%%%%%%%%%%%%%%%%%%%%%%%%%%%%%
\section{Inflationary Heavy-lifting Mechanism
 \label{sec:heavy-lift}}
%%%%%%%%%%%%%%%%%%%%%%%%%%%%%%%%%%%%%%%%%%%%%

During inflation, we have the following effective field theory (EFT) for the inflaton, denoted by a real scalar $\phi$, and the complex PQ scalar field $\chi$ in the ($+$\;$-$\;$-$\;$-$) convention: 
\begin{align}
 {\cal L} &= (\partial_\mu \phi)^2/2 + |\partial_\mu \chi|^2 - V(\phi, \chi)~, \nonumber \\
	 V(\phi, \chi) & =  V(\phi) + \frac{\lambda}{2} \qty(\abs{\chi}^2 - \frac{f_a^2}{2})^2 + \frac{c \, (\partial \phi)^2}{\Lambda^2} \abs{\chi}^2~, \label{eqn:basicLagrangian}	
\end{align}
where $\Lambda$ denotes the UV cutoff of the EFT, $c$ and $\lambda$ are positive dimensionless Wilson coefficients, and $V(\phi)$ stands for the inflaton potential that drives inflation. Here $\chi$ is the spectator field during inflation, and our discussions below are insensitive to different forms of $V(\phi)$. Our key new ingredient is the dimension-six operator\footnote{There are ways to reduce the dimension of this operator. For instance, similar to Eq.~(5.3) of \cite{Kumar:2017ecc}, one can reduce the dimension of $(\partial \phi) \abs{\chi}^2$ by introducing a new heavy mediator field and two more mass scales. After integrating out the mediator, these scales  and new couplings are lumped into $\Lambda$ and $c$, and do not contribute further to understand the key phenomenology of our mechanism.} in Eq.~\eqref{eqn:basicLagrangian}, which is the leading operator coupling $\phi$ and $\chi$ while respecting the (approximate) shift symmetry of $\phi$. This type of operator has been discussed in completely different contexts, such as cosmological collider~\cite{Kumar:2017ecc, Wang:2020ioa}.\footnote{Operator of this form coupling inflaton to the Higgs field could also lead to imprints on the inflaton spectrum from oscillating electroweak symmetry breaking during inflation ~\cite{Fan:2019udt}.} 
 Following~\cite{Kumar:2017ecc}, we name our mechanism inflationary heavy-lifting of the PQ field.

During inflation, the homogeneous background $\ev{\partial_\mu \phi} = \dot{\phi}_0 \delta_{\mu 0}$ then generates the effective quadratic term of the PQ scalar:
\begin{equation}
	V(\phi, \chi) \supset  \qty(\frac{c \, \dot{\phi}_0^2}{\Lambda^2} - \frac{\lambda}{2} f_a^2) \abs{\chi}^2~.
\end{equation}
Note that the first term $\propto c$ has the opposite sign of the second term. When it is large enough, it can protect the PQ field from breaking during inflation. After the inflation ends, the expectation of $\partial{\phi}$ becomes smaller, allowing the PQ symmetry to be spontaneously broken by the tachyonic mass term. In addition, to reproduce the conventional post-inflationary behavior, the PQ scalar needs to be heavier than the Hubble scale during inflation to settle at the minimum, with its large-scale fluctuations exponentially suppressed. This translates to 
\begin{equation}
	m_{\rm PQ,eff}^{2} \approx \frac{ c\, \dot{\phi}_0^2}{\Lambda^2}  - \frac{\lambda}{2} f_a^2 \gtrsim (1.5 H_I)^2~,
\label{eq:heavyfield}
\end{equation}
where 1.5$H_I$ is the boundary value for a field to be heavy enough in the inflationary background~\cite{Chen:2009zp}. 

We then turn to determine the range of parameters in Eq.~\eqref{eq:heavyfield}. Firstly, the validity of the EFT requires $\Lambda$ to be sufficiently larger than the scale of infrared degrees of freedom. Without referring to a specific UV completion, the heavy-field-induced corrections to inflaton self-interactions generically scale as $\mathcal{O}(\dot{\phi}_0^2/\Lambda^4) (\partial \phi)^2$. Requiring the power expansion in series of $\dot{\phi}_0^2/\Lambda^4$ doesn't spoil the EFT during inflation imposes the condition that $\Lambda\gtrsim \dot{\phi}_0^{1/2}$~\cite{Creminelli:2003iq,Kumar:2017ecc}. Secondly, there is a lower limit of coupling $\lambda$ when the $c$ term is present. Considering the theory to be natural, the radiative correction to $\lambda$, $\delta \lambda$, induced by the inflaton loop shall be subdominant. Integrating out the loop momenta to the cutoff scale $\Lambda$, we have
\begin{equation}
	\delta \lambda \sim \frac{c^2}{16\pi^2} \lesssim \lambda \implies c \lesssim 4\pi \sqrt{\lambda}~.
\label{eq:naturalness}
\end{equation}
Similarly, we require the radiative correction to the quadratic mass term of $\chi$ from the inflaton loop, $\delta m_{\rm PQ}^2$, to satisfy 
\begin{equation}
\delta m_{\rm PQ}^2 \sim \frac{c}{16 \pi^2} \Lambda^2 \lesssim \lambda f_a^2~.
\label{eq:naturalness2}
\end{equation}
Lastly, after inflation, the PQ field will acquire a vacuum expectation value (VEV) and the heavy-lifting operator contributes to the inflaton's kinetic term. To avoid a wrong-sign kinetic term,
\begin{equation}
c f_a^2 \ll \Lambda^2~.
\end{equation}

As a benchmark, we choose $\Lambda = 3  \sqrt{\dot{\phi}_0}$ and $c= \pi \sqrt{\lambda}$. Then Eq.~\eqref{eq:heavyfield} turns into a quadratic relation for $\sqrt{\lambda}$. Requiring that $\lambda$ satisfying this relation is real, we find that 
\begin{equation}
f_a \lesssim \frac{\sqrt{2} \pi}{27} \frac{\dot{\phi}_0}{H_I}~,
\label{eqn:faHI}
\end{equation}
which is saturated when $\sqrt{\lambda} = 81 H_I^2/(2 \pi \dot{\phi}_0)$. One could check that all the requirements listed above are satisfied. It is also worth mentioning that our benchmark is chosen so that the EFT stays away from the boundaries of all the theoretical constraints but not to maximize the $f_a/H_I$ ratio (i.e., a smaller $\Lambda$ and larger $c$ can lead to a larger upper bound on $f_a$ than Eq.~\eqref{eqn:faHI}). Generically, as $\sqrt{\dot{\phi}_0}\gg H_I$ during inflation, PQ symmetry is not broken even for $f_a$ exceeding the conventional upper limit on the PIA, $H_I/(2\pi)$, significantly. Note that $\lambda < c <1$ at the upper bound of $f_a$, which implies that the PQ field has a shallow potential. This could be realized similarly to the supersymmetric saxion scenario in~\cite{Choi:2011rs, Fan:2011ua}. 

The size of $\dot{\phi}_0$ can be estimated by the CMB observations. For single-field slow-roll inflation (case 1), $\dot{\phi}_0$ is constrained by the observed primordial scalar power spectrum $A_s \approx H_I^4/(4\pi^2\dot{\phi}_0^2)$. From the Planck measurement that $A_s\approx 2.1\times10^{-9}$ at large scales~\cite{Planck:2018jri}, we have $\sqrt{\dot{\phi}_0}\approx 60H_I$. From Eq.~\eqref{eqn:faHI}, the viable range for the post-inflationary axion becomes
\begin{equation}
{\rm case\, 1:} \, f_a \lesssim 600 H_I~. 
\label{eq:faHI_noncurvaton}
\end{equation} 

There is an interesting alternative with $\dot{\phi}_0$ determined differently. In this scenario, $A_s$ is generated by another spectator field $\sigma$ with a mass $m_\sigma\ll H_I$ instead of the inflaton. Such a light field $\sigma$ is often referred to as the curvaton~\cite{Enqvist:2001zp,Lyth:2001nq,Moroi:2001ct} (case 2). During inflation, the expected value of $\sigma$ is frozen due to the Hubble friction, while its quantum fluctuations generate $A_s$.\footnote{We do not consider the intermediate scenario with both the curvaton and inflaton contributing to $A_s$ significantly as it is even more constrained~\cite{Smith:2015bln}.} It only dominates the universe much later after the inflation ends and eventually decays back to radiation. The constraint on $\dot{\phi}_0$, in this case, is given by the slow-roll parameter $\epsilon \approx \dot{\phi}_0^2/(2 H_I^2 M_{\rm Pl}^2)$ with the reduced Planck scale $M_{\rm Pl} \approx 2.4 \times 10^{18}$~GeV instead. The upper bound $f_a$ in Eq.~\eqref{eqn:faHI} then relaxes to
\begin{equation}
{\rm case\, 2:} \, f_a \lesssim 8 \times 10^{16} \, {\rm GeV} \sqrt{\frac{\epsilon}{0.02}} ~.
	\label{eqn:CurvatonPhiDot}
\end{equation}
The benchmark of $\epsilon$ is the observational upper bound, determined by the relation that
$n_s -1 \approx -2\epsilon$~\cite{Wands:2002bn,Vennin:2015vfa,Kumar:2019ebj}, with $n_s\approx 0.96$ the scalar spectral index~\cite{Planck:2018jri}.\footnote{Notice that the other constraint $\epsilon \lesssim 0.004$ reported in~\cite{Planck:2018jri} doesn't apply to the curvaton case since the relation $r=16\epsilon$ (from single field inflation) it relies on no longer holds.} 

While we focus on the inflaton derivative coupling, it is also possible to consider non-derivative coupling which still keeps the inflaton shift symmetry breaking under control. As an example, we could consider an operator 
\begin{equation}
\frac{c^\prime V(\phi)}{\Lambda^{\prime 2}} |\chi|^2~,
\end{equation}
where we treat $V(\phi)$ as the inflaton shift symmetry breaking spurion and do not introduce additional symmetry-breaking sources. In general, we could consider a different expansion series of inflaton coupling to the PQ field in powers of $V(\phi)/\Lambda^{\prime 4}$. For this series to hold, $\Lambda^\prime \gtrsim V(\phi)^{1/4}$.\footnote{\label{ftnt:Rchi2}When $\Lambda^\prime \sim M_{\rm pl}$, this term is reduced to the Hubble-induced mass term~\cite{Affleck:1984fy, Bezrukov:2007ep}. In addition, the Hubble-induced mass term in the Einstein frame could be transformed into the non-minimal coupling of the PQ field to gravity, ${\cal R} |\chi|^2$, with the Ricci scalar ${\cal R} = 12 H^2$ during inflation.} The computation for the $f_a$ range is similar to the derivative coupling, with $\dot{\phi}_0$ replaced by $\sqrt{V(\phi)}$. Since $\sqrt{V(\phi)} \sim  \dot{\phi}_0/\sqrt{\epsilon}$, the upper bound on $f_a$ in Eq.~\eqref{eqn:faHI} could be raised by a factor $1/\sqrt{\epsilon}$ ($\sim$ {\cal} {O}(10) for current upper bounds on $\epsilon$).\footnote{We thank Mark Hertzberg for bringing up this possibility.}

%%%%%%%%%%%%%%%%%%%%%%%%%%%%%%%%%%%%%%%%%%%%%%%%%%%%%%%%%%
\section{New Window of Post-inflationary Axion}
\label{sec:newwindow}
%%%%%%%%%%%%%%%%%%%%%%%%%%%%%%%%%%%%%%%%%%%%%%%%%%%%%%%%%%

\begin{figure*} [t!]
\begin{center}
	\includegraphics[width=6.5in]{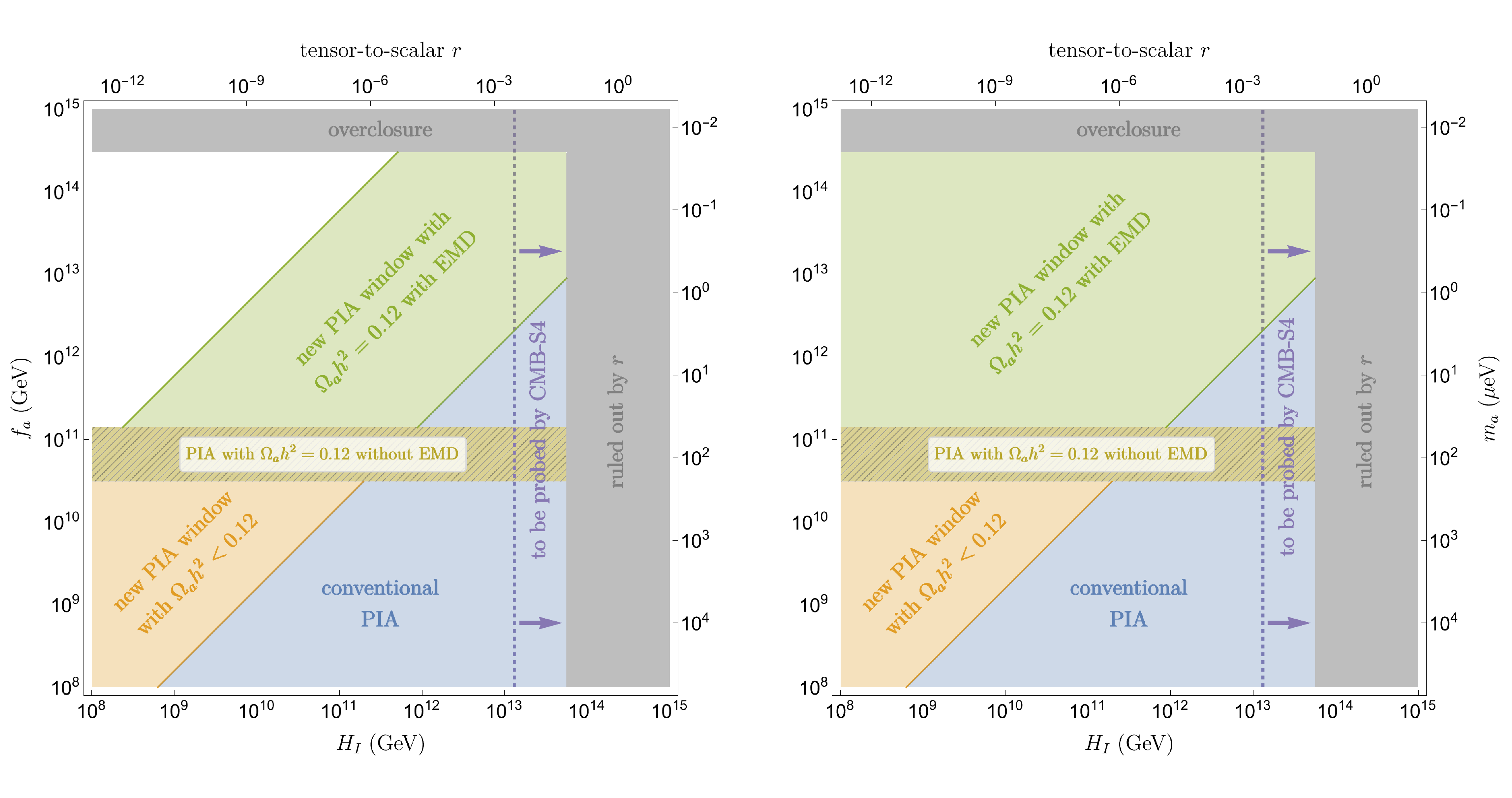}
\end{center}
\caption{New post-inflationary QCD axion window in the ($H_I$,$f_a$) plane. Left: case 1 single-field slow-roll inflation. Right: case 2 the curvaton scenario. Blue: the conventional PIA region with $f_a < H_I/(2\pi)$; orange: new PIA window with axion relic abundance $\Omega_a h^2 < 0.12$; meshed horizontal band: PIA with $\Omega_a h^2 =0.12$ without EMD; green: new PIA window in which $\Omega_a h^2 = 0.12$ with EMD. Region right to the purple dashed line could be probed by CMB-S4~\cite{CMB-S4:2016ple}. The grey regions are either ruled out by the upper bound on the tensor-to-scalar ratio $r$~\cite{Planck:2018jri} (right vertical) or axion DM relic abundance overclosing the universe even with EMD (upper horizontal). }
	\label{fig:new_window}
\end{figure*}

We illustrate how the heavy-lifting mechanism changes the classification of QCD axion cosmologies in~\cref{fig:new_window}. One could see that in the two panels representing case 1 (single-field slow-roll inflation) and case 2 (curvaton scenario) respectively, the upper bounds of $f_a$ based on Eq.~\eqref{eq:faHI_noncurvaton} and Eq.~\eqref{eqn:CurvatonPhiDot} extend the conventional PIA region (blue) substantially. In case 2, the entire space in the $(H_I, f_a)$ plane could belong to PIA.

However, with a high $f_a$, the post-inflationary scenario often overproduces axion DM via the radiation and decay of the axion string network. For the case that the universe is purely radiation dominated between inflaton reheating and matter-radiation equality, only when $f_a\sim 10^{10-11}$ GeV the axion string network can produce the observed DM relic abundance $\Omega_{\rm DM} h^2\approx 0.12$~\cite{Gorghetto:2020qws,Buschmann:2021sdq}. Here we take the numerical benchmark of $f_a\in [3.1, 14]\times 10^{10}$~GeV~\cite{Buschmann:2021sdq}, represented as the meshed horizontal bands in~\cref{fig:new_window}. PIA is still allowed when $f_a$ is lower, but additional sources of DM are needed (shown as orange regions in~\cref{fig:new_window}). More parameter space with a higher $f_a$ is accessible if the DM abundance can be modified in a non-standard cosmology. In this work, we focus on the EMD scenario, where the universe's energy density is dominated by some non-relativistic particles after inflation reheating~\cite{Lazarides:1990xp,Kawasaki:1995vt, Banks:1996ea, Giudice:2000ex,Grin:2007yg}.\footnote{Though non-minimal, EMD is highly motivated theoretically, e.g., it could be sourced by the moduli arising ubiquitously from string theory constructions~\cite{Moroi:1994rs, Kane:2015jia} or dark glueballs~\cite{Foster:2022ajl}.} In this case, the axion first oscillates during the EMD at a time $t^\ast$, which satisfies $3H(t^\ast)\approx m_a(t^\ast)~$\cite{Kawasaki:1995vt,Nelson:2018via}.\footnote{In this case, $t^\ast$ could be very different from the one when the QCD axion starts to oscillate during the radiation domination epoch.} After that, the massive particles causing the EMD decay at the time scale of their lifetime $\sim \Gamma_{\rm EMD}^{-1}$. The universe undergoes a secondary reheating as the energy density of the massive particles is converted into radiation, which dominates afterward. The entropy created by massive particles greatly dilutes the axion number density per comoving entropy density. The axion relic abundance $\Omega_a h^2$ is thus suppressed. Due to tight constraints, the EMD has to end before the Big Bang nucleosynthesis (BBN) era: $\Gamma_{\rm EMD}^{-1}$ must be shorter than the onset of BBN around $\mathcal{O}(1)$~second~\cite{Kawasaki:2017bqm}. Equivalently, the radiation temperature right after the EMD ends, $T_R$, must be larger than $\mathcal{O}(1)$~MeV. When $T_R > \SI{1}{\MeV}$, $\Omega_{a}h^2=0.12$ can be satisfied when $f_a \lesssim \SI{2.7E14}{\GeV}$. For a conservative range of $T_R > \SI{10}{\MeV}$, a large $f_a$ of ${\cal O} (10^{13} - 10^{14})$ GeV is still allowed. More detailed computations can be found in~\cref{app:EMDRelicAbund}. The new parameter regime satisfying $\Omega_{a}h^2=0.12$ after introducing the EMD dilution mechanism with $T_R > \SI{1}{\MeV}$ for the inflaton (curvaton) case is shown in green in \cref{fig:new_window}.

One potential concern is that the EMD-inducing particles may have its own independent fluctuations as curvatons, resulting in additional isocurvature constrains. This is not always true, at least in case 1. They could be generated by the thermal bath created from the inflaton reheating (see~\cite{Dondi:2019olm,Asadi:2022vkc} for examples) so inheriting the inflaton's perturbations; or they could have time-dependent potentials with much larger masses during inflation and thus have much suppressed quantum fluctuations~\cite{Dine:1995kz, Iliesiu:2013rqa}. Then in case 1, the density perturbations originate from a single source, the inflaton perturbations. Then there is no isocurvature constraint on the PIA in the new window.

In case 2, the curvaton scenario, the situation is more subtle. Now the curvaton $\sigma$ induces the EMD. As a spectator field during inflation, the curvaton's quantum fluctuation $\delta \sigma$ is not correlated with the inflaton's fluctuation. Such differences can again introduce isocurvature perturbations between radiation, baryon, neutrino, and DM. Consequently, the isocurvature bounds could still be stringent if the PIA DM from the string networks obtains its perturbations from early radiation before the EMD~\cite{Smith:2015bln}. While the axion string network is created much before EMD, it does not behave as the regular cold DM with a fixed comoving density. In contrast, it is well known that the comoving string density follows the attractive solution of the scaling law, which is insensitive to the initial conditions~\cite{Kibble:1976sj,Kibble:1980mv,Vilenkin:1981kz,Fleury:2015aca,Klaer:2017qhr,Gorghetto:2018myk}. In other words, the large-scale correlation carried by the axion string network in the early universe could be washed out during the long EMD era. More discussions on axion dynamics with an EMD era can be found in the appendix.

For other observational constraints, several cavity searches have already probed the QCD axion within $m_a \subset (1.9, 24.0)$ $\mu$eV, or correspondingly $f_a \subset (2.5 - 32) \times 10^{11}$ GeV (not the full range; only discrete subsets) through the axion-photon coupling, assuming QCD axion is the entire DM~\cite{ADMX:2003rdr,Brubaker:2016ktl, HAYSTAC:2018rwy, ADMX:2018gho,ADMX:2019uok,Lee:2020cfj, HAYSTAC:2020kwv, Jeong:2020cwz, CAPP:2020utb, ADMX:2021nhd, Yoon:2022gzp, Kim:2022hmg}. More parameter space in \cref{fig:new_window} is still open to be explored.

%%%%%%%%%%%%%%%%%%%%%%%%%%%%%%%%%%%%%%%%%%%%%%%%%%%%%%%%%%
\section{Conclusions}
\label{sec:conc}
%%%%%%%%%%%%%%%%%%%%%%%%%%%%%%%%%%%%%%%%%%%%%%%%%%%%%%%%%%

In this article, we propose an inflationary heavy-lifting mechanism to make the PQ field heavy during inflation, and thus the PQ symmetry remains unbroken even when $f_a \gg H_I$. This could be achieved by including the leading high-dimensional operator coupling the inflaton to the PQ field, which is inflaton shift symmetric. This simple inclusion drastically modifies the landscape of the QCD axion cosmologies. The post-inflationary QCD axion parameter space expands significantly in the $(H_I, f_a)$ plane. The region subject to the axion isocurvature constraint shrinks in a correlated way, alleviating the tension between the QCD axion DM and high-scale inflation. Other types of operators, which keep the inflaton shift symmetry breaking under control, also exist to achieve the same goal. We also apply EMD to make more parameter space of high $f_a$ consistent with the observed DM abundance.  

We will outline several interesting consequences and possible future directions: {\it 1)} we describe our mechanism using an EFT. It could be useful to construct the UV completions and examine whether there could be other cosmological observables beyond $r$, such as primordial non-Gaussianities; {\it 2)} our proposal demonstrates that the intriguing late-universe phenomenologies of PIA, such as the formation of miniclusters, could happen in a much larger territory of the inflation-axion parameter space than considered before, and thus calls for more studies; {\it 3)} While we focus on the QCD axion, the mechanism could also be applied to axion-like particles in general and expand their post-inflationary regimes as well; {\it 4)} beyond the well-motivated EMD, one could consider other mechanisms (e.g.~\cite{Allali:2022yvx}) to open up more parameter space for PIA DM of high $f_a$ without overclosing the universe, and their associated model-dependent signals.

\section*{Acknowledgments}
We thank Keisuke Harigaya, Mark Hertzberg, Soubhik Kumar, Matt Reece, and Yi Wang for enlightening discussions and comments on the draft.   
JF and LL are supported by the DOE grant DE-SC-0010010 and the NASA grant 80NSSC22K081.

\appendix 

\section{Axion relic abundance in early matter domination
 \label{app:EMDRelicAbund}}

In this appendix, we first review the calculation of $\Omega_a h^2$ from misalignment mechanism without EMD~\cite{Preskill:1982cy,Dine:1982ah,Abbott:1982af}. During the radiation domination, the cosmic evolution of the homogeneous axion background is governed by the equation of motion:
\begin{equation}
\ddot{a} + 3H(T) \dot{a} + m_a(T)^2 a = 0,
\end{equation}
in which $H(T)$ denotes the Hubble scale at the temperature $T$. In the early universe, we have $H \gg m_a$ so that the axion field value is frozen due to the Hubble friction. However, as the radiation temperature drops, $m_a(T)$ increases while $H(T)$ decreases. Eventually, at the axion oscillating temperature $T_*$ around $3H(T_*) \approx m_a(T_*)$, the axion potential becomes important, and the axion starts to oscillate about its minimum. The axion energy density at $T_*$ can be approximated by 
\begin{equation}
\rho_a(T_*) \approx \frac{1}{2} m_a(T_*)^2 a_i^2 =\frac{1}{2} m_a(T_*)^2 f_a^2 \theta_i^2~,
\label{eqn:adensity}
\end{equation}
in which the dimensionless angle $\theta \equiv a / f_a$ and the subscripts $i$ denote the initial values of the quantities. In the discussions above, we assume that the axion potential is harmonic, which is not exact. However, as long as $\theta_i$ is not very close to $\pi$, the harmonic potential is still a good numerical approximation with only mild corrections~\cite{Turner:1985si,Lyth:1991ub,Strobl:1994wk}. With the approximate harmonic potential, the comoving axion number density is an adiabatic invariant when $T<T_*$:
\begin{equation}
\label{eqn:axiondensityTstar}
R(T)^3 n_a(T) \approx R(T)^3 \frac{\rho_a(T)}{m_a(T)}\approx  \frac{R(T_*)^3}{2}m_a(T_*)f_a^2 \theta_i^2~,
\end{equation}
where $R(T)$ is the scale factor at temperature $T$ and $n_a(T)$ is the axion number density. It is then easy to estimate $\Omega_a$ since $\Omega_a\equiv \rho_a(T_0)/\rho_{\rm crit} = m_a(T_0)n_a(T_0)/\rho_{\rm crit}$, where $T_0$ is today's radiation temperature and $\rho_{\rm crit}$ is the current critical energy density of the universe.

The result above changes after a phase of EMD is induced by a massive particle $\psi$. In particular, if the axion starts to oscillate before the EMD ends, $\Omega_a$ will be diluted by the entropy generated during the reheating. We first estimate the new oscillation temperature $T_*$ given the reheating temperature $T_R$. Here $T_R$ is defined by the radiation temperature immediately after the massive particle decays. It is well known that during the EMD, the temperature of the radiation bath scales like $T \propto t^{-1} \propto  \rho_\psi^{1/8}$~\cite{Steinhardt:1983ia}. Then, using the approximation that $H(T_*) \approx m_a(T_*)/3$ and the Friedmann equation, we have~\cite{Kawasaki:1995vt}
\begin{align}
3H(T_*)^2 \MPl^2 &\approx \rho_\psi(T_R) \frac{T_*^8}{T_R^8} \\\nonumber  \implies 
m_a(T_*) \MPl T_R^2 &\approx \sqrt{\frac{\pi^2 g_*(T_R)}{10}} T_*^4,
\label{eqn:EMDTOsc}
\end{align}
where $g_*(T)$ is the effective degrees of freedom at $T$ and we use $\rho_\psi(T_R) = \pi^2 g_*(T_R)T_R^4/30$.

After the reheating of EMD, $\Omega_a$ can be estimated via
\begin{align}
\Omega_a  & \approx \frac{1}{\rho_{\rm crit}} m_a(T_0) \frac{\rho_a(T_R)}{m_a(T_R)}  \frac{s(T_0)}{s(T_R)} ~,
\end{align}
where $s(T)$ is the entropy density at $T$. The relation above uses the fact that after EMD ends, $n_a/s$ is conserved. The axion energy density at $T_R$ is estimated by
\begin{align}
\label{eqn:EMDTOsc2}
\rho_a(T_R) =  \rho_a(T_*) \frac{m_a(T_R)}{m_a(T_*)} \frac{\rho_\psi(T_R)}{\rho_\psi(T_*)}~. 
\end{align}
In Eq.~\eqref{eqn:EMDTOsc2} we used the fact that $n_a(T)/n_\psi(T) = \text{const.}$ during EMD. Combining Eqs.~\eqref{eqn:adensity},~\eqref{eqn:EMDTOsc},  and~\eqref{eqn:EMDTOsc2}, one finds that the axion relic density generated by the misalignment mechanism during EMD is 
\begin{equation}
\Omega_a^{\rm mis} h^2 = \frac{h^2}{2\rho_{\rm crit}} m_a(T_*)m_a(T_0) f_a^2 \theta_i^2  \frac{h_*(T_0)}{h_*(T_R)} \frac{T_0^3 T_R^5}{T_*^8}~,
\label{eqn:axionMisalignRelic}
\end{equation}
where $h_*(T)$ is the effective degrees of freedom that contributes to the entropy at $T$ and $h \approx 0.7$.

To estimate the temperature dependence of $m_a(T_*)$, we use the result extrapolated from lattice simulations~\cite{Dine:2017swf}
\begin{equation}
m_a(T) = 
\begin{dcases}
	\frac{\sqrt{\chi(0)}}{f_a}, & T<T_2, \\
	\frac{\sqrt{\chi(T_0)}}{f_a} \qty(\frac{T_0}{T})^{n/2}, & T_2<T<T_0, \\
	\frac{\sqrt{\chi(T_0)}}{f_a} \qty(\frac{T_0}{T})^4, & T>T_0,
\end{dcases}
\label{eqn:TDptMa}
\end{equation}
in which the zero-temperature QCD susceptibility $\chi(0) = \SI{3.6E-5}{\GeV^4}$, the demarcating temperature for the high-temperature regime $T_0 = \SI{1.5}{\GeV}$, the high-temperature susceptibility $\chi(T_0) = \SI{3.7E-14}{\GeV^4}$, and the intermediate-temperature benchmark $T_2 = \qty(\chi(T_0) / \chi(0))^{1/n} T_0$ which makes $\chi(T)$ continuous. The intermediate power dependence $n$ is a number ranging from $7$ to $20$. Combining Eq.~\eqref{eqn:EMDTOsc},~\eqref{eqn:EMDTOsc2},~ \eqref{eqn:axionMisalignRelic}, and~\eqref{eqn:TDptMa}, one can obtain the relic density of the QCD axion DM from misalignment mechanism as a function of $f_a$ and $T_R$. For PIA, the effective $\theta_i^2$ takes the average value $\langle\theta_i^2\rangle= \pi^2/3$.

However, the dominant contribution to $\Omega_a$ in the PIA case is the radiation and decay of the axion string network, which could be much larger than $\Omega_a^{\rm mis}$. The numerical simulation of the axion relic density contributed from the string network $\Omega_a^{\rm str}$ is known to be difficult. Here we take the simplified approach  that the ratio between $\Omega_a^{\text{str}}$ and $\Omega_a^{\text{mis}}$ is a fixed constant $\alpha>1$~\cite{Nelson:2018via}. Recent simulation \cite{Buschmann:2021sdq} estimates that $\alpha \sim 6$ to $8$. Using $\alpha \sim 8$, we find that for $T_R = \SI{1}{\MeV}$, the constraint that $\Omega_a^{\text{mis}} + \Omega_a^\text{str} = \Omega_\text{DM}$ provides an upper bound on $f_a \lesssim \SI{2.7E14}{\GeV}$, which is insensitive to the choice of $n$ in Eq.~\eqref{eqn:TDptMa}. 
On the other hand, when $T_R = \SI{10}{\MeV}$, the parameter $n$ could be a dominant source of uncertainty. Nonetheless, one still find that the upper bound for the PQ scale lands on $\sim {\cal O} (10^{13} - 10^{14})$ GeV for $T_R = \SI{10}{\MeV}$.

For the complicated axion string dynamics, multiple uncertainties may be present in our calculation. For example, the axion string density may scale with time differently from the one extrapolated in lattice simulations as the physical time span greatly exceeds the capability of simulations~\cite{Hindmarsh:2021zkt}. It is also possible that dimensionless parameters describing the string network evolution differ from our reference values by $\mathcal{O}(1)$ factors~\cite{Gouttenoire:2019kij}, as most lattice simulations assume a radiation-dominated background cosmology. Therefore, the upper bound on $f_a$ derived when $T_{\rm R}\approx 1$~MeV is a suggestive value, while a dedicated lattice simulation during EMD is needed for more precise predictions. However, most newly opened-up parameter space pointed out by this work is still valid aside from the part very close to the overclosure line.

Finally, we would like to comment briefly on other possible sources of isocurvature in the curvaton scenario with PIA DM. In Sec.~\ref{sec:newwindow}, we focus on the isocurvature from axion string decays. We argue that the attractor solution of the string network washes out the initial-condition dependence of the PIA profile. The fluctuation of PIA relic density is mainly due to that of the curvaton. Other components of PIA DM either come from axion string radiation before the string decays or the misalignment mechanism. However, PIA DM from string radiation is relativistic and redshift as radiation; thus, this contribution is expected to be minor and dominated by late-time contributions during the EMD without significant isocurvature~\cite{Gorghetto:2018myk}. On the other hand, for the misalignment mechanism contribution to the PIA DM abundance, the misalignment angle is randomized over $[-\pi,\pi)$. Hence, no large-scale isocurvature is generated like the standard PIA in a radiation domination epoch. Only small-scale fluctuations, such as axion minihalos, can be produced and observed. A dedicated numerical study of this scenario is needed in the future to validate the arguments above fully. 
 \\ \\

\bibliography{Ref}
\end{document}